# Hydrogen Bonding: A Mechanism for Tuning Electronic and Optical Properties of Hybrid Organic – Inorganic Frameworks

Fedwa El-Mellouhi[1], El Tayeb Bentria[1], Asma Marzouk[1], Sergey N. Rashkeev[1*],
Sabre Kais[1,2,3], and Fahhad H. Alharbi[1,2]

[1]Qatar Environment & Energy Research Institute (QEERI), Hamad Bin Khalifa University, P. O. Box 5825, Doha, Qatar

[2]College of Science and Engineering, Hamad Bin Khalifa University, Doha, Qatar

[3]Department of Chemistry and Physics, Birck Nanotechnology Center, Purdue University, West Lafayette, IN 47907, USA

*email: srashkeev@qf.org.qa

**Here we demonstrate that significant progress in this area may be achieved by introducing structural elements that form hydrogen bonds with environment. Considering several examples of hybrid framework materials with different structural ordering containing protonated sulfonium cation $H_3S^+$ that forms strong hydrogen bonds with electronegative halogen anions ($Cl^-$, $F^-$), we found that hydrogen bonding increases the structural stability of the material and may be used for tuning electronic states near the bandgap. We suggest that such a behavior has a universal character and should be observed in hybrid inorganic-organic framework materials containing protonated cations. This effect may serve as a viable route for optoelectronic and photovoltaic applications.**

In the past few years, hybrid organic–inorganic frameworks have aroused great interest due to the rich variety of their physical and chemical properties and potential application in nanotechnologies. These systems are prepared by binding multi-dentate organic molecules to multi-coordinate metal complexes and provide unlimited combinations of structures and properties, which enables to target design of materials for many applications.[1] In some cases, binding may also be achieved and/or assisted by other mechanisms such as halogen bonding[2,3] or hydrogen bonding.[4] One of the most important issues is controlling and tuning the structural, optical, thermal, mechanical, and electronic properties of these complex materials by varying their chemistry, fabrication techniques, and preparation conditions. On one side, porous hybrid framework materials for applications in chemical sensing,[5] gas storage,[6] and catalysis[7] were developed. On the other side, there is a great interest to develop dense hybrid metal-organic systems for applications in optical and semiconductor devices and batteries.[8] Some of dense hybrid frameworks such as [AmH]M(HCOO)$_3$ adopt the ABX$_3$ perovskite architecture (A = protonated amine, [AmH]$^+$, B = M$^{2+}$ is a divalent metal, and X = HCOO$^-$ is the formate anion) and exhibit wide range of ferroelectric and multiferroic properties.[9,10]

Another class of hybrid perovskites of the composition [AmH]MX$_3$ (M = Sn$^{2+}$, Pb$^{2+}$ and X = I$^-$, Br$^-$, Cl$^-$) which were known for more than two decades,[11] has recently attracted a great deal of interest. In particular, lead based iodides MAPbI$_3$ and FAPbI$_3$ (MA = CH$_3$NH$_3^+$ and FA = CH(NH$_2$)$_2^+$ are methylammonium and formamidinium ions) show outstanding performance in solar cell applications and may be synthesized by straightforward processing methods such as spin-coating, dip-coating, and vapor deposition techniques.[12-14] The electrical power conversion efficiency of perovskite devices has shot up dramatically and now reached 21.1%,[15] which is believed to be the result of unique combination of the orbital character of the conduction and valence band extrema, large absorption coefficient and long carrier diffusion lengths which result in low recombination rates.[16,17] It was also suggested that material stability and defect tolerance in these perovskites as well as in other organic-inorganic frameworks may be related to the presence of van der Waals interactions and hydrogen bonding.[18-20] Recent quantum molecular dynamics simulations indicated that iodine atoms form hydrogen bonds with H atoms located at the nearest MA$^+$ ion, i.e., charge fluctuations due to continuous formation and breakage of these bonds significantly contribute to the dielectric function of MAPbI$_3$.[21]

Motivated by these latest discoveries, we predict the possibility of tuning electronic properties of hybrid organic-inorganic materials and stabilizing them by hydrogen bonds, or electrostatic binding of hydrogen to a nearby highly electronegative atom (N, O, F, etc.). This type of bonding occurs in both inorganic (water) and organic (DNA and proteins) molecules and plays an important role in determining the three-dimensional structures adopted by these molecules such as double helical DNA structure. Many polymers (e.g., nylon), are also strengthened by hydrogen bonds in their main chains. Here, we are considering several groups of hybrid organic-inorganic materials in which hydrogen bonds play a significant role in: (i) tuning the band structure of the material (e.g., the orbital character of electronic bands near the bandgap) by forming new bonds and changing hybridization between the orbitals, and; (ii) stabilizing the framework structure to be more resistant to structural transformations. The considered groups of organic-inorganic materials are: (i) The three-dimensional (3D) tetragonal perovskite structures (β- phase) of H$_3$SPbX$_3$ (X = I$^-$, Br$^-$, Cl$^-$, F$^-$) containing the sulfonium cation H$_3$S$^+$; (ii) The structural δ- phase of the same materials containing a planar two-dimensional (2D) structure formed by PbX$_6$ octahedra with common edges in their equatorial planes, and; (iii) The 2D material with experimentally known RbPbF$_3$ structure in which Rb$^+$ cation is substituted by a protonated sulfonium H$_3$S$^+$ cation which forms hydrogen bonds with fluorine.

We performed first-principles density functional theory (DFT) based calculations (see details in the Methods section) for crystalline unit cells of these three groups of materials and studied the character of wave functions of the states near the bandgap. For this purpose, we projected the density of states (DOS) onto atomic orbitals of different angular momenta (*s*-, *p*-, and *d*-) centered at different atoms. Figure 1 shows the tetragonal perovskite structure (β-phase) of H$_3$SPbI$_3$ (Fig. 1a), the band structure of H$_3$SPbCl$_3$ (Fig. 1b), and DOSes of four perovskite materials, H$_3$SPbX$_3$ (X = I$^-$, Br$^-$, Cl$^-$, F$^-$) projected onto different atomic species (Figs. 1c, d, e, f). Analysis of the projected DOS in the vicinity of the bandgap indicates that sulfonium cation makes a significant difference in the orbital character of states near the bandgap as compared to MAPbI$_3$ and FAPbI$_3$. In MAPbI$_3$, the electronic band gap is formed between the antibonding top of the valence band (VB) originating from the Pb(6*s*)–I(5*p*) interactions and the antibonding conduction band (CB) minimum resulting from the Pb(6*p*)–I(5*p*) interactions. The MA$^+$ cation does not introduce any states at the band edges (the partial contributions from C, N, and H atomic orbitals at the energy interval of 5 eV below the VB maximum do not exceed 10$^{-3}$ of the total DOS, Ref.[22]) , which could be the reason of high tolerance of this material to some defects.[17]

In the H$_3$SPbX$_3$ perovskites, the value of the bandgap is mainly defined by the size of the anion (the smallest for F$^-$ and largest for I$^-$) and its electronegativity (the lowest for I$^-$ and highest for F$^-$)[23] (Supporting Information, Table S1) which is similar to the MAPbX$_3$ materials. However, in the H$_3$SPbX$_3$ perovskites one could observe much higher contributions of the sulfur (S) and hydrogen (H) states at the energy range of 0.5 – 4 eV below the top of the VB, which means that sulfonium cation orbitals are hybridizing with the halogen anion X(*p*) orbitals which provide the major contribution to the DOS in this area. This hybridization is higher for anions with higher electronegativity (or smaller anion radius) being the largest for H$_3$SPbF$_3$ and lowest for H$_3$SPbI$_3$ (Figs. 1c, d, e, f). The hybridized H$_3$S-X states form additional features at the DOS curves – one or more noticeable shoulders or even local peaks near the top of the VB. An analysis based on the projected DOS onto atomic orbitals of different angular momenta centered at different atoms (Figure S1) shows that these features are related to the hybridization of S(3*p*) – H(1*s*) – X(*p*) orbitals which increases energy distances between the upper valence bands (Fig. 1b).

In the structural δ- phases of H$_3$SPbX$_3$ materials which crystallize in unit cell four times larger than the perovskite unit cell, the planar 2D structure is formed by PbX$_6$ octahedra that share edges in their equatorial planes (Figure S2a). We found that qualitative behavior of δ- modifications of these materials is very similar to the behavior of their 3D perovskite counterparts, i.e., the δ- phase DOSes projected onto different atomic species also indicate a significant admixture of the S and H orbitals of the sulfonium ions near the top of the VB (Figure S2). The bandgaps for the 2D δ- phase materials containing the I⁻, Br⁻, and Cl⁻ ions are higher than in same materials in the 3D α- phase while for $H_3SPbF_3$, the bandgaps are similar for both 2D and 3D phases (Table S1).

We also performed calculations for the 2D material $H_3SPbF_3$ that crystallizes in experimentally known $RbPbF_3$ structure, in which $Rb^+$ cations are substituted by protonated $H_3S^+$ cations. The structure of this material is different from the δ- phase structures considered above – here the 2D planes consist of $PbF_4$ tetrahedra (instead of octahedra in the δ- phase) that have common edges (Fig. 2a). We found that $H_3S$ molecules in this framework also form hydrogen bonds with fluorine anions, and the electronic structure near the top of VB is similar to the two other groups of materials considered earlier – in addition to the fluorine p- states, there is a strong admixture of the S(3$p$) and H(1$s$) states that form two extra peaks in the DOS near the top of VB (Fig. 2b). Figure 2c shows that these peaks should be observed in the imaginary part of the dielectric function, $ε_2(ω)$, at UV energies around 4.5 and 5.3 eV for any possible polarization of the electric field (i.e., they will be also observed in polycrystalline samples and/or in unpolarized light experiments).

All of these results support the assumption that protonated sulfonium cation $H_3S^+$ interacts with halogen anion, the interaction being stronger for anions with higher electronegativity (Cl⁻ and F⁻). To understand the nature of this interaction, we plotted the electronic localization function (ELF) contours[24] for several considered materials in the vicinity of the $H_3S$-X complexes. In perovskite structure, the ELF minimum between the $H_3S$ complex and the iodine atom corresponds to the absence of bonding between them while an S-H-F electronic bridge is formed with fluorine (Figs. 3a and 3b). This is a typical example of hydrogen bonding in which hydrogen atom electrostatically attracts to strongly electronegative fluorine. This hydrogen bond has some features of covalent bond as well – it is directional and forms interatomic distances shorter than the sum of the van der Waals radii. An even stronger H-F hydrogen bond is observed in the $H_3SPbF_3$ material with $RbPbF_3$ structure (Fig. 3c) which means that S-H-F bridges could be observed in many inorganic-organic framework containing sulfonium cations and highly electronegative halide anions.

The high S-H-F angle of 167 – 177° observed in all considered materials indicates that there is competition between the S-H covalent bond and hydrogen attraction to electronegative anion (Cl or F). Most likely, such a behavior has even more universal character and could be observed in materials containing protonated cations based on atoms with electronegativity lower than nitrogen (e.g., phosphorus). The length of the N-H covalent bond (1.04 Å) is much lower than of the S-H and P-H bonds (1.34 and 1.44 Å) which is an important geometric factor explaining why protonated amines (e.g., MA and FA, for which most of previous studies on inorganic-organic photovoltaic devices have been performed) prefer not to form hydrogen bonds with surrounding atoms even with those with high electronegativity. In sulfates and phosphates, however, this restriction is lifted which provides an opportunity to form extra hydrogen bonds which could: (i) contribute into forming the unique structures of the network; (ii) increase their stability, and; (iii) provide fine tuning of the states at the edges on the bandgap.

Our calculations indicate that considerable structures exhibit reasonable stability. In addition to the formation energy, we calculated the reaction (difference between the total energy of a reaction and reactants)[25] and the Hull (the difference in formation energies which evaluates the stability of a given compound against any linear combination of stable phases)[26] energies for β- and δ- phases (Tables S2 and S3). The results for $H_3SPbX_3$ (X = I, Br, Cl, and F) materials show that all of these energies are negative except of the reaction energy for the β- phase of $H_3SPbF_3$ which is zero within the accuracy of the calculations.

Here we demonstrated that significant progress in tuning electronic and optical properties of hybrid organic – inorganic frameworks may be achieved by introducing structural elements that form hydrogen bonds with environment. Considering several examples of hybrid framework materials with different structural ordering containing protonated sulfonium cation $H_3S^+$ that forms strong hydrogen bonds with electronegative halogen anions ($Cl^-$, $F^-$), we found that hydrogen bonding increases the structural stability of the material and may be used for tuning electronic states near the bandgap. We suggest that such a behavior has a universal character and should be observed in different hybrid inorganic-organic framework materials containing protonated cations. This effect may serve as a viable route for optoelectronic and photovoltaic applications. The potential tunability of the states near the edges of the band gap (VB maximum and CB minimum) could also provide an important unique tool for developing different devices. Hydrogen bonding mixes up the states of the protonated cation and the anion which gives a contribution to DOS and to the optical absorption spectra at specific frequencies close to the bandgap threshold (peaks or shoulders) which could be identified and selectively excited by laser. This could offer a new approach to different devices with promising applications in nanoelectronics, optics, and data storage.

**Methods**

We employ first principles calculations to evaluate the electronic structure and to estimate the stability of the proposed materials. DFT calculations are performed with the projector augmented wave (PAW) method as implemented in the Vienna Ab-initio Simulation Package (VASP).[27] All calculations were performed using spin polarized generalized gradient approximation (GGA) with the Perdew−Burke−Ernzerhof (PBE) parametrization for the exchange and correlation energy of interacting electrons.[28] The energy cutoff for the planewave basis set was set to 520 eV and a 8×8×8 Monkhorst−Pack *k*-point mesh was employed. Long-range van der Waals interactions have been taken into account via the Tkatchenko and Scheffler (TS) scheme[29]. The convergence of the final forces is set to 0.01 eV/Å.

**Acknowledgements**

All authors designed the research. FEM performed calculations. All participated in data analysis and writing the paper. The simulations were performed at the Research Computing Center in Texas A&M University at Qatar and the Swiss Supercomputing Center (CSCS).

**Additional Information**

Supporting Information is available from the Wiley Online Library or from the author.

**Competing financial interests**: The authors declare no competing financial interests.

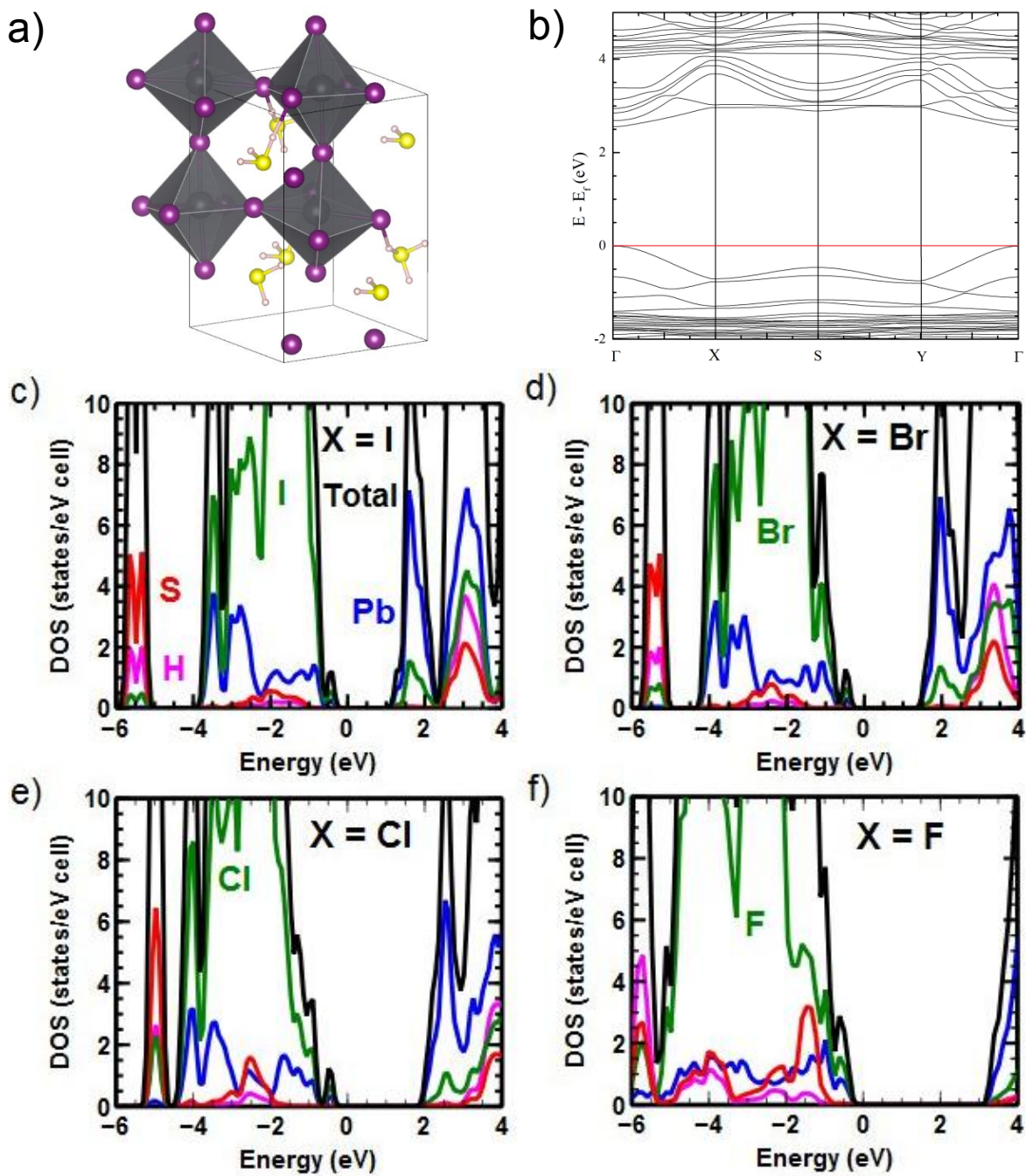

**Figure 1 | Crystal structure and electronic properties of perovskite $H_3SPbX_3$ materials. a,** The schematic of periodic unit cell of $H_3SPbI_3$ tetragonal perovskite (β- phase). Pb atoms are shown in gray, I – in purple, S – in yellow, H – in white. **b,** Electronic band structure of tetragonal $H_3SPbCl_3$. Zero of energy is positioned at the top of the VB. **c, d, e,** and **f**, Density of states (total and projected at different atomic species) for β- phase of $H_3SPbI_3$, $H_3SPbBr_3$, $H_3SPbCl_3$, and $H_3SPbF_3$, respectively. Black line indicates the total DOS; green, blue, red, and magenta – partial contributions of halogen (X), lead (Pb), sulfur (S), and hydrogen (H).

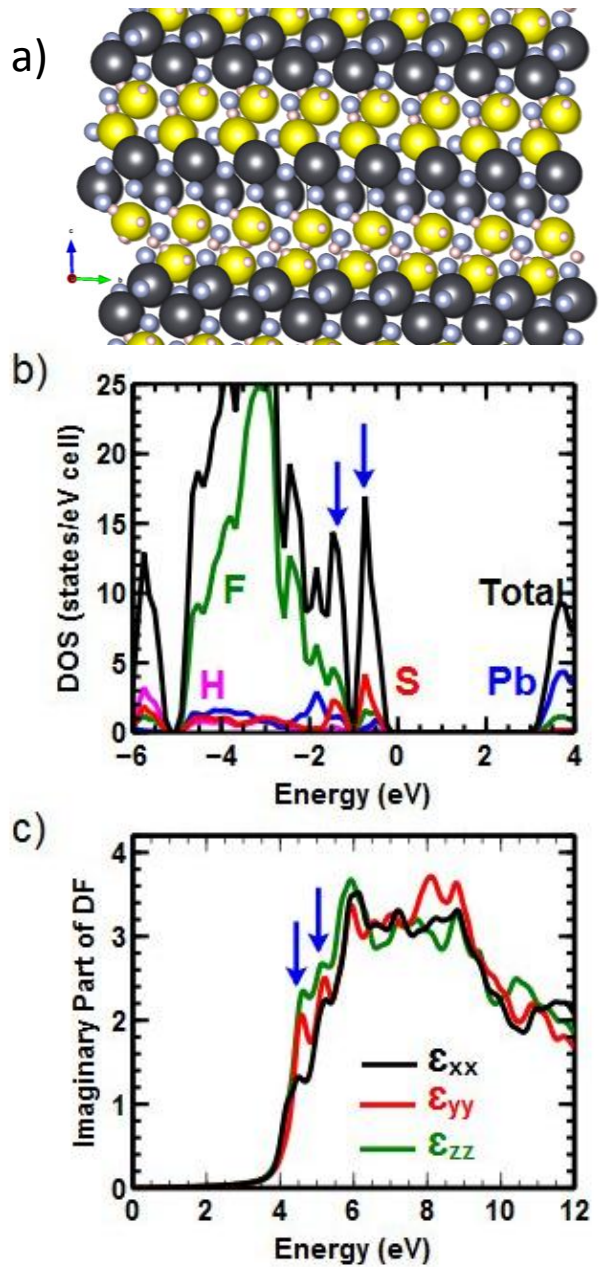

**Figure 2 | Electronic properties of $H_3SPbF_3$ material crystallizing in the $RbPbF_3$ structure. a,** The schematic of $H_3SPbF_3$ crystal structure in the space filling mode. Pb atoms are shown in dark gray, F – in light gray, S – in yellow, H – in white. **b**, Electronic density of states (total and projected at different atomic species) of $H_3SPbF_3$, material crystallizing in $RbPbF_3$ structure. Black line indicates the total DOS; green, blue, red, and magenta – partial contributions of fluorine (F), lead (Pb), sulfur (S), and hydrogen (H). **c,** Calculated imaginary part of dielectric function (DF) for this material for three different polarizations of the electric field. Vertical blue arrows indicate interband electronic transitions from electronic states positioned near the top of VB (also indicated with blue arrows in **b**) to the edge of the conduction band.

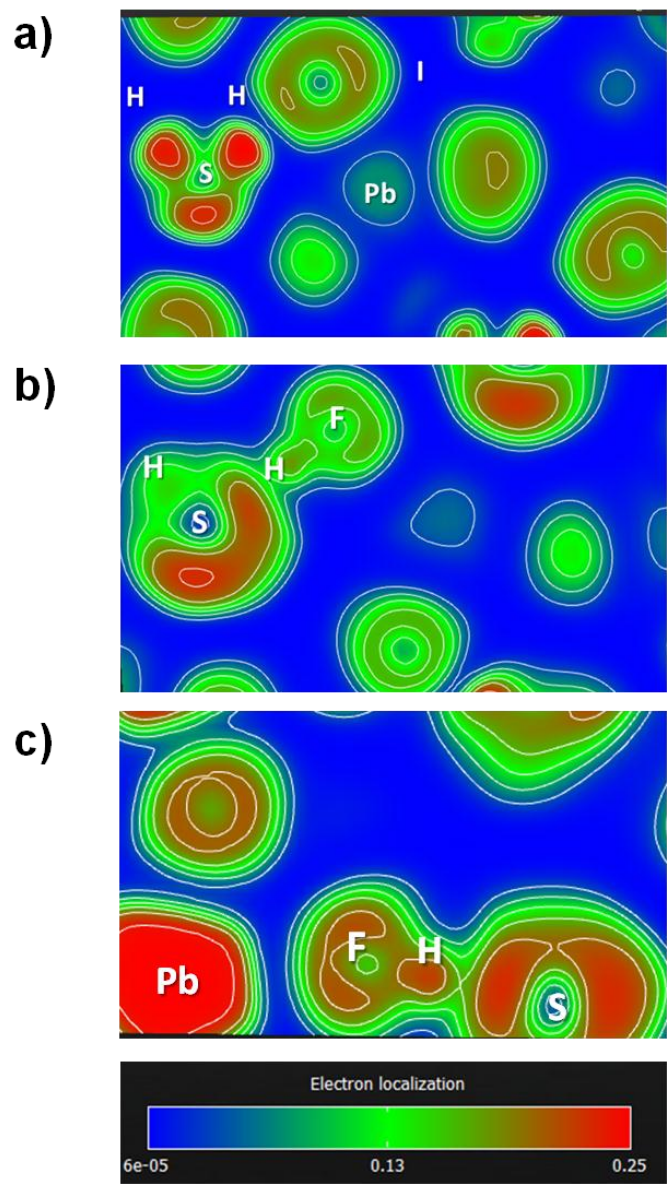

**Figure 3 | ELF contours of different H$_3$SPbX$_3$ materials in the vicinity of H$_3$S-X complexes. a,** ELF for H$_3$SPbI$_3$ in perovskite structure (β- phase). **b**, ELF for H$_3$SPbF$_3$ in perovskite structure. **c,** ELF for H$_3$SPbF$_3$ crystallizing in RbPbF$_3$ structure.

# Supporting Information

**Hydrogen Bonding: A Mechanism for Tuning Electronic and Optical Properties of Hybrid Organic – Inorganic Frameworks**

Fedwa El-Mellouhi[1], El Tayeb Bentria[1], Asma Marzouk[1], Sergey N. Rashkeev[1*], Sabre Kais[1,2,3], and Fahhad H. Alharbi[1,2]

[1]Qatar Environment & Energy Research Institute (QEERI), Hamad Bin Khalifa University, P. O. Box 5825, Doha, Qatar

[2]College of Science and Engineering, Hamad Bin Khalifa University, Doha, Qatar

[3]Department of Chemistry, Birck Nanotechnology Center, Purdue University, West Lafayette, IN 47907, USA

*email: srashkeev@qf.org.qa

**Table S1 | Calculated electronic bandgaps for $H_3SPbX_3$ (X = I, Br, Cl, and F) materials in β- and δ-phases.**

| Halogen Atom (X) | I | Br | Cl | F |
|---|---|---|---|---|
| Bandgap (eV) in β- phase (3D) | 1.73 | 2.06 | 2.5 | 3.78 |
| Bandgap (eV) in δ- phase (2D) | 2.49 | 2.88 | 3.44 | 3.76 |

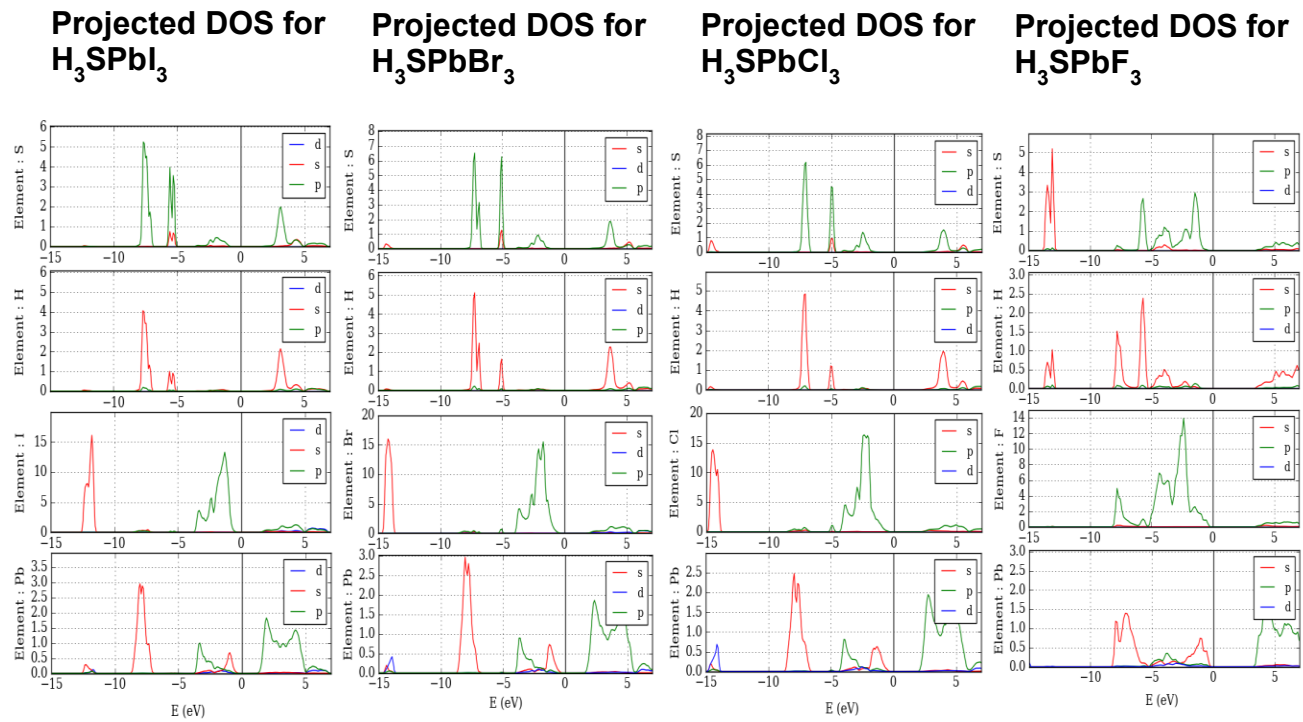

**Figure S1 | DOS for H$_3$SPbX$_3$ (X = I, Br, Cl, and F) materials in β- phase projected at different atomic species and different angular momentum (*s*-, *p*-, and *d*-) orbitals.** Contribution of *s*- orbitals is shown by red lines, *p*- orbitals – by green lines, and *d*- orbitals – by blue lines.

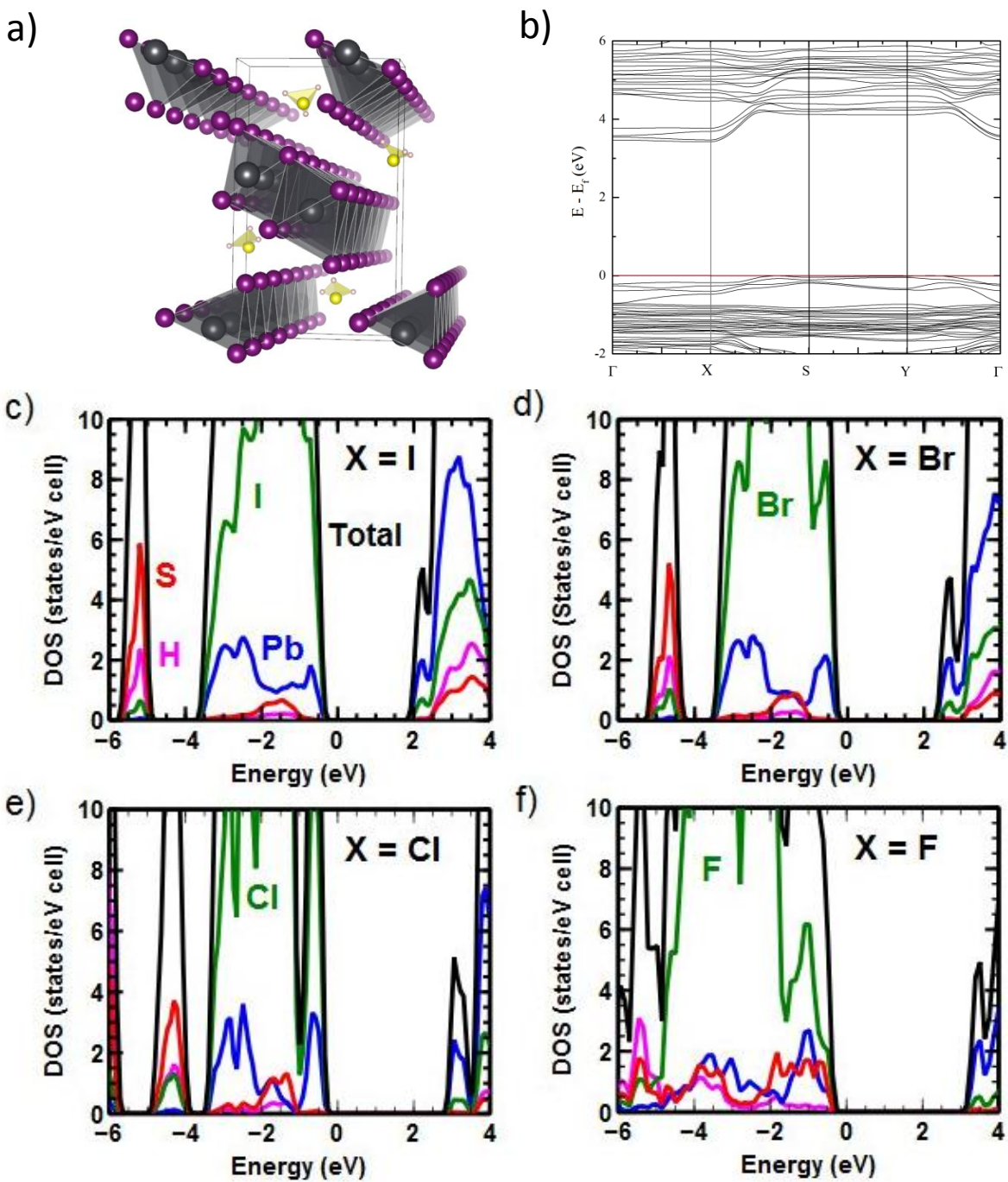

**Figure S2 | Crystal structure and electronic properties of $H_3SPbX_3$ materials in δ- phase. a,** The schematic of the crystal structure of δ- phase of $H_3SPbI_3$. The periodic unit cell is indicated by thin lines. Pb atoms are shown in gray, I – in purple, S – in yellow, H – in white. **b**, Electronic band structure of $H_3SPbCl_3$ in δ- phase. Zero of energy is positioned at the top of the VB. **c, d, e,** and **f**, Density of states (total and projected at different atomic species) for δ- phase of $H_3SPbI_3$, $H_3SPbBr_3$, $H_3SPbCl_3$, and $H_3SPbF_3$, respectively. Black line indicates for total DOS; green, blue, red, and magenta – partial contributions of halogen (X), lead (Pb), sulfur (S), and hydrogen (H).

**Table S2 | Calculated reaction, formation and Hull energies for $H_3SPbX_3$ (X = I, Br, Cl, and F) materials in β- phase (in eV/cell).**

| Material | Reaction Energy* | Formation Energy | Hull Energy |
|---|---|---|---|
| $H_3SPbI_3$ | -0.21 | -0.48 | -0.30 |
| $H_3SPbBr_3$ | -0.16 | -0.63 | -0.25 |
| $H_3SPbCl_3$ | -0.10 | -0.97 | -0.18 |
| $H_3SPbF_3$ | 0.01 | -1.55 | -0.07 |

**Table S3 | Calculated reaction, formation and Hull energies for $H_3SPbX_3$ (X = I, Br, Cl, and F) materials in δ- phase (in eV/cell).**

| Material | Reaction Energy* | Formation Energy | Hull Energy |
|---|---|---|---|
| $H_3SPbI_3$ | -0.18 | -0.44 | -0.26 |
| $H_3SPbBr_3$ | -0.14 | -0.61 | -0.23 |
| $H_3SPbCl_3$ | -0.08 | -0.96 | -0.17 |
| $H_3SPbF_3$ | -0.02 | -1.58 | -0.28 |

*Reaction Energies for $H_3SPbX_3$ (X = I, Br, Cl, and F) materials were calculated for the reactions: $PbS + 3HX \rightarrow H_3SPbX_3$